\documentclass[twocolumn,5p,number,letterpaper]{elsarticle}%
\usepackage{amsmath}
\usepackage{epsfig}
\usepackage{natbib}
\usepackage{accents}
\usepackage{amsfonts}
\usepackage{amssymb}
\usepackage{graphicx}%
\providecommand{\U}[1]{\protect\rule{.1in}{.1in}}
\pdfoutput=1
\providecommand{\U}[1]{\protect\rule{.1in}{.1in}}

\journal{Solid State Communications}
\begin{document}
%
%TCIMACRO{\TeXButton{frontmatter}{\begin{frontmatter}
%\title{Thermoelectric spin diffusion in a ferromagnetic metal}
%\author{Moosa Hatami\fnref{label1}}
%				
%\author{Gerrit E. W. Bauer\fnref{label1}}
%\fnref{label1}
%\author{Saburo Takahashi\fnref{label2}}
%\author{Sadamichi Maekawa\fnref{label2,label3}}
%\address
%[label1]{Kavli Institute of NanoScience, Delft University of Technology,
%Lorentzweg 1, 2628 CJ Delft, The Netherlands}
%\address[label2]{Institute for Materials Research, Tohoku University, Sendai,
%Miyagi 980-8577, Japan}
%\address
%[label3]{CREST, Japan Science and Technology Agency, Tokyo 100-0075, Japan}
%\begin{abstract}
%We present a semiclassical theory of spin-diffusion in a ferromagnetic metal
%subject to a temperature gradient.
%Spin-flip scattering can generate pure thermal spin currents
%by short-circuiting spin channels while
%suppressing spin accumulations. A thermally induced spin density
%is locally generated when the energy dependence of the
%density of states is spin polarized.
%\end{abstract}
%\begin{keyword}
%A. Metallic ferromagnets; D. thermoelectrics; D. spin diffusion;  D. spin caloritronics
%\PACS\sep72.15.Jf; 75.60.Jk; p85.75.-d; 75.30.Sg
%\end{keyword}
%\end{frontmatter}}}%
%BeginExpansion
\begin{frontmatter}
\title{Thermoelectric spin diffusion in a ferromagnetic metal}
\author{Moosa Hatami\fnref{label1}}
				
\author{Gerrit E. W. Bauer\fnref{label1}}
\fnref{label1}
\author{Saburo Takahashi\fnref{label2}}
\author{Sadamichi Maekawa\fnref{label2,label3}}
\address
[label1]{Kavli Institute of NanoScience, Delft University of Technology,
Lorentzweg 1, 2628 CJ Delft, The Netherlands}
\address[label2]{Institute for Materials Research, Tohoku University, Sendai,
Miyagi 980-8577, Japan}
\address
[label3]{CREST, Japan Science and Technology Agency, Tokyo 100-0075, Japan}
\begin{abstract}
We present a semiclassical theory of spin-diffusion in a ferromagnetic metal
subject to a temperature gradient.
Spin-flip scattering can generate pure thermal spin currents
by short-circuiting spin channels while
suppressing spin accumulations. A thermally induced spin density
is locally generated when the energy dependence of the
density of states is spin polarized.
\end{abstract}
\end{frontmatter}%
%EndExpansion

\section{Introduction}

Thermoelectric properties attracted new interest in recent years due to the
improved performance of nanometer-scale structures
\cite{Giazotto:rmp06,Sales:sc02,Boukai:nat08,Hochbaum:nat08}. Spin
caloritronics addresses the interplay of conduction electron charge, spin and
entropy/energy transport in solid state nanostructures in the search for
improved heat control. The thermodynamics of spin flow in metallic structures
involving magnetic elements is an important research area. The giant
magneto-thermoelectric power in multilayered nanopillars \cite{Gravier:prb06a}%
, thermally excited spin-currents in metals with embedded ferromagnetic
clusters \cite{Tsyplyatyev:prb06}, thermal spin-transfer torques
\cite{Hatami:prl07} and magneto-Peltier cooling \cite{Hatami:prb09} in
spin-valve devices are examples of spin-dependent thermoelectric phenomena on
a nanometer scale. Our understanding of the coupling between spin and
thermoelectric transport is still very incomplete, however.

Power dissipation is a major problem obstructing further miniaturization of
electronics, thus deserves to be studied in magnetic systems. The Seebeck
effect refers to conversion of a thermal differential into an electric
voltage, for instance in a thermocouple. A spin analogue to the Seebeck effect
has been discovered by Uchida \textit{et al.} \cite{Uchida:nat08} in permalloy
thin films subject to a temperature gradient. The inverse spin Hall voltage
picked up by platinum top contacts is evidence of a thermally induced spin
current locally generated by the ferromagnet, which changes sign in the middle
of the wire and has an approximately linear profile between the heat source
and drain.

Here we discuss temperature-gradient-driven thermoelectric transport in
disordered ferromagnetic metals. Starting from the Boltzmann equation we study
spin and heat diffusion through the ferromagnet in the presence of
spin-conserving and spin-flip impurity scattering. We show that thermal
generation of a spin accumulation is limited by the spin-flip scattering. At
nonzero (finite) temperatures the spin-flip scattering equilibrates the spin
imbalance in the electron distribution function (or spin accumulation), but
not the temperature-dependent magnetization density at local thermal equilibrium.

We start in Sec. 2 by formulating a diffusion theory for a ferromagnetic metal
in the presence of spin-dependent external forces and derive spin diffusion
equations for the distribution functions. Spin dependent thermoelectric
transport is studied in Sec. 3 for a ferromagnetic metal subject to a
temperature gradient. In Sects. 4 and 5 we explain the effect of a temperature
dependence of the chemical potential. The spin densities and exchange
splitting in the presence of temperature-dependent densities of states are
discussed in the Appendix.

\section{Spin Boltzmann and diffusion equations}

We consider a homogeneous ferromagnetic metal in a simple Stoner model with
energies $\varepsilon_{k}^{\left(  \alpha\right)  }=\varepsilon_{k}%
+\alpha\Delta-\mu_{0}\ $for spins $\alpha=\pm1,$ where $\varepsilon_{k}$ is
the kinetic energy for momentum $k=\left\vert \mathbf{k}\right\vert $,
$\Delta\left(  T\right)  $ the temperature dependent exchange splitting, and
$\mu_{0}$ the ground state chemical potential. The Boltzmann equation
describes evolution of the distribution of conduction electrons in the
presence of external fields and random scattering processes and for a steady
state reads
\begin{equation}
\left(  {\mbox{\boldmath$\upsilon$}_{\mathbf{k}}}\cdot
\mbox{\boldmath$\nabla$}_{\mathbf{r}}+\frac{\mathbf{F}^{\left(  \alpha\right)
}}{\hbar}\cdot\mbox{\boldmath$\nabla$}_{\mathbf{k}}\right)  f^{\left(
\alpha\right)  }(\mathbf{k},\mathbf{r})=\left(  \frac{\partial f^{\left(
\alpha\right)  }}{\partial t}\right)  _{\mathrm{scatt.}} \label{Boltzmann1}%
\end{equation}
where ${\mbox{\boldmath$\upsilon$}_{\mathbf{k}}}%
=\mbox{\boldmath$\nabla$}_{\mathbf{k}}\varepsilon_{k}$ is the electron
velocity and $f^{\left(  \alpha\right)  }(\mathbf{k},\mathbf{r})\ $ its
distribution as a function of momentum $\mathbf{k},$ position $\mathbf{r}$ and
spin $\alpha$. The forces $\mathbf{F}^{\left(  \alpha\right)  }\left(
\mathbf{r}\right)  =e\mbox{\boldmath$\nabla$}_{\mathbf{r}}\phi-\alpha\left(
\partial\Delta/\partial T\right)  \mbox{\boldmath$\nabla$}_{\mathbf{r}}T$
originate from the gradients of the electric potential $\phi$ and exchange
splitting $\Delta$, where the latter may include an external (Zeeman) magnetic
field. $e=\left\vert e\right\vert $ is the modulus of the electron charge and
$\hbar=h/2\pi$ Planck's constant. The collision integral on the right-hand
side contains spin-conserving and spin-flip impurity scattering. Inelastic
scattering is assumed to be strong. The isotropic parts of the distribution
functions in momentum space are then locally thermalized to the Fermi-Dirac
form at temperature $T\left(  \mathbf{r}\right)  $ and spin-dependent chemical
potential shifts $\mu^{\left(  \alpha\right)  }\left(  \mathbf{r}\right)  :$
\begin{align}
\bar{f}^{\left(  \alpha\right)  }(\varepsilon_{k},\mathbf{r})  &  =\left(
4\pi\right)  ^{-1}\int d\Omega_{\mathbf{k}}f^{\left(  \alpha\right)
}(\mathbf{k},\mathbf{r})\\
&  =\left[  e^{\left(  \varepsilon_{k}^{\left(  \alpha\right)  }-\mu^{\left(
\alpha\right)  }\left(  \mathbf{r}\right)  \right)  /\left(  k_{B}T\left(
\mathbf{r}\right)  \right)  }+1\right]  ^{-1}.
\end{align}
Inelastic scattering processes then do not appear explicitly in the collision
term of the right-hand side of Eq. (\ref{Boltzmann1}), which in the relaxation
time approximation reads (note the difference with \cite{Hershfield:prb97}):%
\begin{align}
\left(  \frac{\partial f^{\left(  \alpha\right)  }}{\partial t}\right)
_{\mathrm{scatt.}}  &  =-\frac{g^{\left(  \alpha\right)  }(\mathbf{k}%
,\mathbf{r})}{\tau_{\mathrm{tr}}^{\left(  \alpha\right)  }(\varepsilon_{k}%
)}\nonumber\\
&  -\frac{\bar{f}^{\left(  \alpha\right)  }(\varepsilon_{k},\mathbf{r}%
)-\bar{f}^{\left(  -\alpha\right)  }(\varepsilon_{k}+2\alpha\Delta
,\mathbf{r})}{\tau_{\mathrm{sf}}^{\left(  \alpha\right)  }(\varepsilon_{k})}
\label{Boltzmann3}%
\end{align}
where $g^{\left(  \alpha\right)  }(\mathbf{k},\mathbf{r})=f^{\left(
\alpha\right)  }(\mathbf{k},\mathbf{r})-\bar{f}^{\left(  \alpha\right)
}(\varepsilon_{k},\mathbf{r})\ $and $1/\tau_{\mathrm{tr}}^{\left(
\alpha\right)  }(\varepsilon)=1/\tau_{0}^{\left(  \alpha\right)  }%
(\varepsilon)+1/\tau_{\mathrm{sf}}^{\ \left(  \alpha\right)  }(\varepsilon)$
are the relaxation rates that include both the spin-conserving and spin-flip
scattering times $\tau_{0}^{\left(  \alpha\right)  }$ and $\tau_{\mathrm{sf}%
}^{\left(  \alpha\right)  },$ respectively.

Using $\hbar^{-1}\mbox{\boldmath$\nabla$}_{\mathbf{k}}\bar{f}^{\left(
\alpha\right)  }=\mbox{\boldmath$\upsilon$}_{\mathbf{k}}^{\left(
\alpha\right)  }\partial_{\varepsilon_{k}}\bar{f}^{\left(  \alpha\right)  }$
and averaging Eq. (\ref{Boltzmann1}) over momentum directions we find for the
anisotropic part of the electron distribution function
\begin{equation}
g^{\left(  \alpha\right)  }(\mathbf{k},\mathbf{r})=-\tau_{\mathrm{tr}%
}^{\left(  \alpha\right)  }{\mbox{\boldmath$\upsilon$}_{\mathbf{k}}^{\left(
\alpha\right)  }}\cdot\left(  \mathbf{\triangledown}_{\mathbf{r}}%
+\mathbf{F}^{\left(  \alpha\right)  }\frac{\partial}{\partial\varepsilon_{k}%
}\right)  \bar{f}^{\left(  \alpha\right)  }(\varepsilon_{k},\mathbf{r}).
\label{Boltzmann6}%
\end{equation}
Substituting Eq. (\ref{Boltzmann6}) into the Boltzmann Eq. (\ref{Boltzmann1}),
integrating over $\mathbf{k}$-directions yields the spectral equation
\begin{equation}
\mathcal{D}_{k}^{\left(  \alpha\right)  }\left(  \mathbf{\triangledown
}_{\mathbf{r}}+\mathbf{F}^{\left(  \alpha\right)  }\frac{\partial}%
{\partial\varepsilon_{k}}\right)  ^{2}\bar{f}^{\left(  \alpha\right)  }%
=\frac{\bar{f}^{\left(  \alpha\right)  }-\bar{f}^{\left(  -\alpha\right)  }%
}{\tau_{\mathrm{sf}}^{\left(  \alpha\right)  }}. \label{be7}%
\end{equation}
where $\mathcal{D}^{\left(  \alpha\right)  }=%
%TCIMACRO{\TeXButton{upsilon}{\upsilon}}%
%BeginExpansion
\upsilon
%EndExpansion
_{k}^{2}\tau_{\mathrm{tr}}^{\left(  \alpha\right)  }/3$ is the diffusion
constant. Introducing the spectral current density
\begin{equation}
\mathbf{j}^{\left(  \alpha\right)  }(\varepsilon,\mathbf{r})=\frac{1}{e}%
\sigma^{\left(  \alpha\right)  }(\varepsilon)\left(  \mathbf{\triangledown
}_{\mathbf{r}}+\mathbf{F}^{\left(  \alpha\right)  }\frac{\partial}%
{\partial\varepsilon}\right)  \bar{f}^{\left(  \alpha\right)  }(\varepsilon
,\mathbf{r}),
\end{equation}
the charge current
\begin{equation}
\mathbf{J}^{\left(  \alpha\right)  }(\mathbf{r})\equiv-e\int\frac{d\mathbf{k}%
}{(2\pi)^{3}}\mbox{\boldmath$\upsilon$}_{\mathbf{k}}^{\left(  \alpha\right)
}f^{\left(  \alpha\right)  }(\mathbf{k},\mathbf{r})=\int d\varepsilon
\mathbf{j}^{\left(  \alpha\right)  }(\varepsilon,\mathbf{r})\ \label{jc}%
\end{equation}
and the energy current%
\begin{equation}
\mathbf{J}_{E}^{\left(  \alpha\right)  }(\mathbf{r})=-\frac{1}{e}\int
d\varepsilon\varepsilon\mathbf{j}^{\left(  \alpha\right)  }(\varepsilon
,\mathbf{r}) \label{je}%
\end{equation}
are expressed in terms of the conductivity $\sigma^{\left(  \alpha\right)
}=e^{2}\mathcal{N}^{\left(  \alpha\right)  }\mathcal{D}^{\left(
\alpha\right)  }$, in which $\mathcal{N}^{\left(  \alpha\right)  }$ is the
density of states. The charge and energy currents must be conserved,
\textit{i.e.} $\sum_{\left(  \alpha\right)  }%
\mbox{\boldmath$\nabla$}_{\mathbf{r}}\cdot\mathbf{J}^{\left(  \alpha\right)
}(\mathbf{r})=0$ and $\sum_{\left(  \alpha\right)  }%
\mbox{\boldmath$\nabla$}_{\mathbf{r}}\cdot\mathbf{J}_{E}^{\left(
\alpha\right)  }(\mathbf{r})=0.$

The problem is simplified by linearizing Eq. (\ref{be7}). We disregard the
higher order term $\mathbf{F}^{\left(  \alpha\right)  }g^{\left(
\alpha\right)  }$ and the temperature dependence of the material parameters.
Expanding $\bar{f}^{\left(  \alpha\right)  }$ around the equilibrium
distribution function $f_{0}^{\left(  \alpha\right)  },$ we obtain for the
local distribution function in the spin-flip terms
\begin{equation}
\bar{f}^{\left(  \alpha\right)  }(\varepsilon_{k},\mathbf{r}%
)|_{T=\mathrm{const.}}\approx f_{0}^{\left(  \alpha\right)  }(\varepsilon
_{k})+\left(  -\frac{\partial f_{0}^{\left(  \alpha\right)  }}{\partial
\varepsilon_{k}}\right)  \mu^{\left(  \alpha\right)  }\left(  \mathbf{r}%
\right)  ,
\end{equation}
and for the divergence%
\begin{align}
\mathbf{\triangledown}_{\mathbf{r}}\bar{f}^{\left(  \alpha\right)
}(\varepsilon_{k},\mathbf{r})  &  \approx\left(  -\frac{\partial
f_{0}^{\left(  \alpha\right)  }}{\partial\varepsilon_{k}}\right) \nonumber\\
&  \left[  \mathbf{\triangledown}_{\mathbf{r}}\mu^{\left(  \alpha\right)
}+\left(  \frac{\varepsilon_{k}^{\left(  \alpha\right)  }}{T}-\alpha
\frac{\partial\Delta}{\partial T}\right)  \mathbf{\triangledown}_{\mathbf{r}%
}T\right]
\end{align}
After integration over momenta and invoking the Sommerfeld expansion, we
obtain the particle diffusion equation%
\begin{equation}
\nabla_{\mathbf{r}}^{2}\left(  \tilde{\mu}^{\left(  \alpha\right)
}-eS^{\left(  \alpha\right)  }T\right)  =\frac{\mu^{\left(  \alpha\right)
}-\mu^{\left(  -\alpha\right)  }}{\left(  l_{\mathrm{sf}}^{\left(
\alpha\right)  }\right)  ^{2}}, \label{chempots}%
\end{equation}
where $l_{\mathrm{sf}}^{\left(  \alpha\right)  }=\sqrt{\mathcal{D}^{\left(
\alpha\right)  }\tau_{\mathrm{sf}}^{\left(  \alpha\right)  }},$ $\tilde{\mu
}^{\left(  \alpha\right)  }=\mu^{\left(  \alpha\right)  }-e\phi,$ $\ $and the
spin-dependent Seebeck coefficients $S^{\left(  \alpha\right)  }%
=-e\mathcal{L}_{0}T\mathcal{\partial}_{\varepsilon}\ln\sigma^{\left(
\alpha\right)  }|_{\mu_{0}}$ (Mott's formula) in terms of the Lorenz constant
$\mathcal{L}_{0}=\left(  \pi^{2}/3\right)  \left(  k_{B}/e\right)  ^{2}$ have
been introduced. By multiplying Eq. (\ref{be7}) by $\varepsilon_{k}^{\left(
\alpha\right)  }$ and integrating, we obtain the heat diffusion equation%
\begin{equation}
\nabla_{\mathbf{r}}^{2}\left(  -S^{\left(  \alpha\right)  }\tilde{\mu
}^{\left(  \alpha\right)  }+e\mathcal{L}_{0}T\right)  =S_{\mathrm{sf}}%
\frac{\mu^{\left(  \alpha\right)  }-\mu^{\left(  -\alpha\right)  }}{\left(
l_{\mathrm{sf}}^{\left(  \alpha\right)  }\right)  ^{2}}, \label{heat}%
\end{equation}
where due to the detailed balance of spin-flip scattering $S_{\mathrm{sf}%
}=-e\mathcal{L}_{0}T\mathcal{\partial}_{\varepsilon}\ln\left(  N^{\left(
\pm\alpha\right)  }/\tau_{\mathrm{sf}}^{\left(  \pm\alpha\right)  }\right)
|_{\mu_{0}}$ does not depend on spin.

From Eqs. (\ref{chempots}) and (\ref{heat})%
\begin{equation}
\nabla_{\mathbf{r}}^{2}\left(  \frac{S^{2}}{\mathcal{L}_{0}}\left(
1-P^{2}\right)  P_{S}\mu_{s}+\left(  1+\frac{S^{2}}{\mathcal{L}_{0}}\right)
T\right)  =0,
\end{equation}
where $\sigma=\sigma^{\left(  \uparrow\right)  }+\sigma^{\left(
\downarrow\right)  },\;P=\left(  \sigma^{\left(  \uparrow\right)  }%
-\sigma^{\left(  \downarrow\right)  }\right)  /\sigma,$ $\sigma S=\sigma
^{\left(  \uparrow\right)  }S^{\left(  \uparrow\right)  }+\sigma^{\left(
\downarrow\right)  }S^{\left(  \downarrow\right)  },$ $P_{S}=\left(
S^{\left(  \uparrow\right)  }-S^{\left(  \downarrow\right)  }\right)  /S,$ and
$\mu_{s}=\mu^{\left(  \uparrow\right)  }-\mu^{\left(  \downarrow\right)  }.$
For metals the dimensionless parameter $S^{2}/\mathcal{L}_{0}$ $\ll1$. By
letting $S^{2}/\mathcal{L}_{0}\rightarrow0$ the temperature gradient becomes
constant and the conventional diffusion equations for the spin $\mu_{s}$ and
charge accumulation $\tilde{\mu}_{c}=\left(  \tilde{\mu}^{\left(
\uparrow\right)  }+\tilde{\mu}^{\left(  \downarrow\right)  }\right)  /2$ are
recovered \cite{Hershfield:prb97}:%
\begin{align}
\nabla_{\mathbf{r}}^{2}\mu_{s}\left(  \mathbf{r}\right)  -\mu_{s}\left(
\mathbf{r}\right)  /l_{\mathrm{sf}}^{2}  &  =0,\\
\nabla_{\mathbf{r}}^{2}\left[  \tilde{\mu}_{c}\left(  \mathbf{r}\right)
+P\mu_{s}\left(  \mathbf{r}\right)  /2\right]   &  =0,
\end{align}
with $l_{\mathrm{sf}}=\left(  \left(  l_{\mathrm{sf}}^{\left(  \uparrow
\right)  }\right)  ^{-2}+\left(  l_{\mathrm{sf}}^{\left(  \downarrow\right)
}\right)  ^{-2}\right)  ^{-1/2}$.$\ $

Let us consider a conducting magnetic wire with length $\Lambda$ in the $x$
direction. When we fix the charge and spin distribution functions at the
left(right) ends to be $\mu_{cL(R)}$ and $\mu_{sL(R)}$, respectively,
\begin{align}
\mu_{c}(x)  &  =\frac{1}{2}\left(  \mu_{cL}+\mu_{cR}\right)  +\frac{P}%
{4}\left(  \mu_{sL}+\mu_{sR}\right) \nonumber\\
&  \;\;\;\;\;\;+\left[  (\mu_{cR}-\mu_{cL})+\frac{P}{2}(\mu_{sR}-\mu
_{sL})\right]  \frac{x}{\Lambda}-\frac{P}{2}\mu_{s}(x),\label{muc}\\
\mu_{s}(x)  &  =\mu_{sL}\frac{\sinh\left(  \lambda/2-x/l_{_{sf}}\right)
}{\sinh\lambda}+\mu_{sR}\frac{\sinh\left(  \lambda/2+x/l_{_{sf}}\right)
}{\sinh\lambda}, \label{mus}%
\end{align}
where $\lambda=\Lambda/l_{\mathrm{sf}}$ measures the spin-flip scattering.
When $\lambda\gg1$ and $\left\vert x/l_{\mathrm{sf}}-\lambda\right\vert \gg$
$1$ we find that $\mu_{s}(x)=e^{\lambda/2}\left(  \mu_{sL}e^{-x/l_{\mathrm{sf}%
}}+\mu_{sR}e^{x/l_{\mathrm{sf}}}\right)  /\sinh\lambda$ is appreciable only in
proximity of the edges. The vanishing of $\mu_{s}$ in the bulk is independent
of the thermoelectric forces applied by a temperature or chemical potential
difference over the wire. By iteratively reintroducing the heat generation by
spin-flip processes that are of order $S^{2}/\mathcal{L}_{0}$ we observe that
their effect is confined to the edges where $\nabla_{\mathbf{r}}^{2}\mu_{s}$
is significant. Gradients of the exchange potential and external Zeeman
magnetic fields do not appear explicitly at all. We conclude that the spin
accumulation cannot persist over distances longer than the spin-flip diffusion
length. Metallic spin diffusion therefore cannot explain the observed
spin-Seebeck signal in permalloy with $L$/$l_{\mathrm{sf}}\sim10^{5}$
\cite{Uchida:nat08}.

\section{Thermoelectric transport and thermal spin currents}

The local Fermi-Dirac spin distribution functions are defined by spatially and
spin-dependent chemical potentials. Temperatures depend on position, but are
not spin-dependent in the presence of the sufficiently strong inelastic
scattering assumed here. In this Section we consider transport of the
conduction electron spins in a magnetic metal experiencing thermoelectric
forces, \textit{viz}. an external electric field and gradients of the local
chemical potentials and temperature. Using the Sommerfeld expansion in Eqs.
(\ref{jc},\ref{je}) we obtain for the local spin particle and heat currents
$\mathbf{\dot{Q}}^{\left(  \alpha\right)  }=\mathbf{J}_{E}^{\left(
\alpha\right)  }(\mathbf{r})-\mu_{0}\mathbf{J}^{\left(  \alpha\right)
}(\varepsilon,\mathbf{r})$, valid for $\mathcal{L}_{0}T^{2}\left\vert
\partial_{\varepsilon}^{2}\sigma\left(  \varepsilon\right)  |_{\varepsilon
_{F}}\right\vert \ll\sigma\left(  \varepsilon_{F}\right)  $,%
\begin{equation}
\left(
\begin{array}
[c]{c}%
\mathbf{J}^{\left(  \alpha\right)  }\\
\mathbf{\dot{Q}}^{\left(  \alpha\right)  }%
\end{array}
\right)  =\sigma^{\left(  \alpha\right)  }\left(
\begin{array}
[c]{cc}%
1 & S^{\left(  \alpha\right)  }\\
S^{\left(  \alpha\right)  }T & \mathcal{L}_{0}T
\end{array}
\right)  \left(
\begin{array}
[c]{c}%
\mbox{\boldmath$\nabla$}_{\mathbf{r}}\tilde{\mu}^{\left(  \alpha\right)  }/e\\
-\mbox{\boldmath$\nabla$}_{\mathbf{r}}T
\end{array}
\right)  . \label{currents2}%
\end{equation}
The spin-dependent thermal conductivities obey the Wiedemann-Franz law
$\kappa^{\left(  \alpha\right)  }\approx\mathcal{L}_{0}T\sigma^{\left(
\alpha\right)  }$ in the limit $S^{\uparrow(\downarrow)}\ll\sqrt
{\mathcal{L}_{0}}$ and the total thermal conductivity $\kappa=\kappa
^{\uparrow}+\kappa^{\downarrow}=\mathcal{L}_{0}T\sigma$. Eq. (\ref{currents2})
can be rewritten in terms of the charge, spin and heat currents,
$\mathbf{J}_{c(s)}=\mathbf{J}^{\left(  \uparrow\right)  }\pm\mathbf{J}%
^{\left(  \downarrow\right)  }$ and $\mathbf{\dot{Q}}=\mathbf{\dot{Q}%
}^{\left(  \uparrow\right)  }+\mathbf{\dot{Q}}^{\left(  \downarrow\right)  },$
respectively,
\begin{equation}
\left(
\begin{array}
[c]{c}%
\mathbf{J}_{c}\\
\mathbf{J}_{s}\\
\mathbf{\dot{Q}}%
\end{array}
\right)  =\sigma\left(
\begin{array}
[c]{ccc}%
1 & P & S\\
P & 1 & P^{\prime}S\\
ST & P^{\prime}ST & \mathcal{L}_{0}T
\end{array}
\right)  \left(
\begin{array}
[c]{c}%
\mbox{\boldmath$\nabla$}_{\mathbf{r}}\tilde{\mu}_{c}/e\\
\mbox{\boldmath$\nabla$}_{\mathbf{r}}\mu_{s}/2e\\
-\mbox{\boldmath$\nabla$}_{\mathbf{r}}T
\end{array}
\right)  ,
\end{equation}
in which $P$ and $P^{\prime}$ stand for the spin-polarization of conductivity
($\sigma^{\left(  \alpha\right)  }|_{\varepsilon_{F}}$) and its energy
derivative ($\partial_{\varepsilon}\sigma^{\left(  \alpha\right)
}|_{\varepsilon_{F}}$).

In the following we concentrate on the expressions for charge and spin
chemical potential distributions induced in a one-dimensional magnetic wire by
a constant temperature gradient, $\Delta T/\Lambda,$ which is part of an open
electric circuit, \textit{i.e.} in the absence of a charge current. The three
unknowns $\mu_{sL(R)}$and $\Delta\mu$ are determined by boundary conditions.%

\begin{figure}[ptb]%
\centering
\includegraphics[
%trim=0.000000in 0.499150in 0.000000in 0.500217in,
%natheight=3.555200in,
%natwidth=3.555200in,
%height=7.6689cm,
%width=8.0367cm
% added by arXiv admin:
width=\columnwidth
]%
{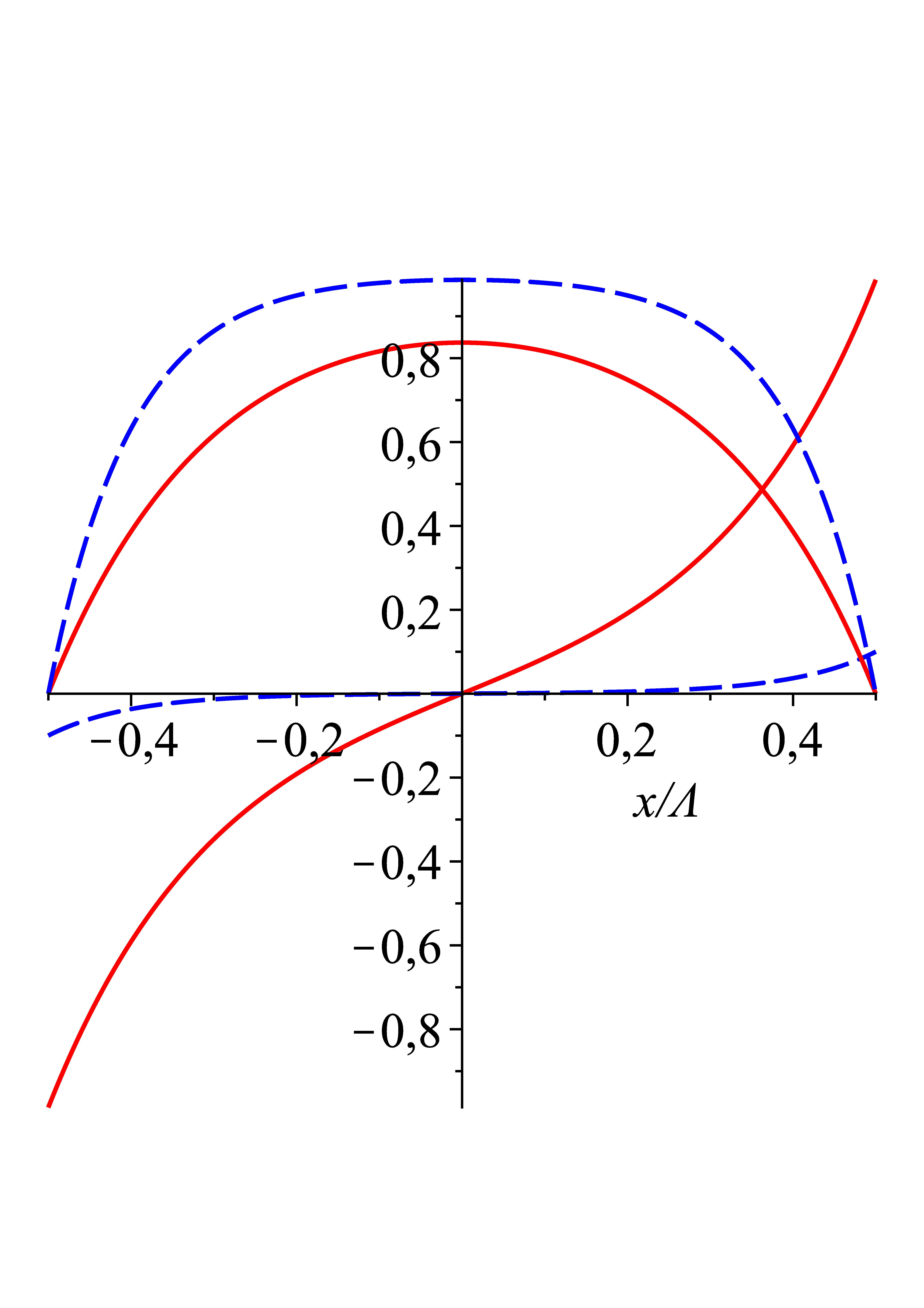}%
\caption{Spatial variation of the thermally induced spin accumulation (changes
sign) and spin current in a magnetic wire for $\lambda=\Lambda/l_{sf}=5$
(solid line) and $10$. The scaling factor of the spin accumulation is
$eP_{S}S\Delta T/\left(  1+PP_{S}\right)  $ and of the spin current
$\sigma(P-P^{\prime})S\Delta T/\Lambda.$}%
\label{Figure1}%
\end{figure}

The chemical potential and temperature differences over the wire are defined
as $\Delta\tilde{\mu}=\mu_{R}-\mu_{L}$, $\Delta T=T_{R}-T_{L}$. The open
electric circuit boundary condition is $J_{c}(x)=0$. We consider two
additional limiting boundary conditions: (i) the reservoirs at the two ends of
the wire are efficient spin sinks so that $\mu_{sL(R)}=0$; (ii) spin are not
dissipated at the ends of the wire whatsoever, \textit{i.e}. $J_{s}%
(-\Lambda/2)=J_{s}(\Lambda/2)=0$. The reality might be somewhere between (i)
and (ii). The heat current $\dot{Q}$ is uniform to leading order in
$S^{2}/{\mathcal{L}}_{0}$.

In the presence of ideal boundary spin sinks (i) the spin accumulation, Eq.
(\ref{mus}), vanishes everywhere in the sample. The total chemical potential,
Eq. (\ref{muc}), changes linearly in the sample and we find $\Delta\tilde{\mu
}=eS\Delta T$, which implies a uniform thermally excited spin current
$J_{s}=(P^{\prime}-P)\sigma S\Delta T/\Lambda$ when $P^{\prime}\neq P$ or
equivalently $P_{S}\neq0$. This situation is similar to the classical
thermocouple in which the wires of different materials ($S_{A}\neq S_{B}$) are
electrically short-circuited at the ends.

Without boundary spin sinks, (ii) $J_{s}(0)=J_{s}(\Lambda)=0$, the spin
accumulation distribution in the wire, Eq. (\ref{mus}), becomes%
\begin{equation}
\mu_{s}\left(  x\right)  =\frac{eP_{S}S\Delta T}{\left(  1+PP_{S}\right)
\lambda}\frac{\sinh\left(  x/l_{_{sf}}\right)  }{\cosh\left(  \lambda
/2\right)  }, \label{musx}%
\end{equation}
changes sign when crossing the center, $\mu_{s}(0)=0$, and its modulus
increases exponentially at the edges (see Fig. \ref{Figure1}). At the ends
$x=\pm\Lambda/2$ we find finite spin accumulations
\begin{equation}
\mu_{sL}=-\mu_{sR}=-\frac{eP_{S}S\Delta T}{\left(  1+PP_{S}\right)  \lambda
}\tanh\frac{\lambda}{2}.
\end{equation}
Also this has an analogue with the classical thermopower effect
\cite{Uchida:nat08}: a thermocouple consisting of wires from two different
materials, creates a voltage difference in the absence of currents
($J_{A(B)}=0$). The spin accumulation has an effect of the measurable
thermopower $S_{c}=\Delta\tilde{\mu}/\left(  e\Delta T\right)  =\left(
S+\mu_{sR}/e\right)  \Delta T$. The spin-flip scattering suppresses spin
accumulation, but generates a thermally induced spin current%

\begin{equation}
J_{s}=\frac{\sigma\left(  P-P^{\prime}\right)  S\Delta T}{\Lambda}\left(
1-\frac{\sinh\left(  x/l_{_{sf}}\right)  }{\cosh\left(  \lambda/2\right)
}\right)  ,
\end{equation}
analogous to the electric current in a short-circuited thermocouple. When
$\lambda\gg1$ the spin current equals that in (i) except at the edges, where
it exponentially decays. In the opposite limit, in the absence of spin-flip
scattering, the spin accumulation adopts a linear relation between contacts
$\mu_{s}(x)\propto x\Delta T/\Lambda$ and the spin current vanishes.

For efficient spin sinks at the edges (i), the spin current is uniform but the
spin accumulation vanishes. The thermally induced spin accumulation in case
(ii) is also suppressed sufficiently far away from the terminals. The spin
accumulation is proportional to the temperature gradient and does not depend
on length of the wire. A spin Seebeck effect should be observable in
ferromagnetic wires in proximity of the edges. However, our results do not
explain the (approximately) linear inverse spin Hall voltage profile measured
by Uchida\textit{\ et al}. \cite{Uchida:nat08} in a permalloy thin film.

\section{Temperature dependence of chemical potential and spin density}

In this Section we discuss the temperature dependence of the chemical
potential and spin density. The strong screening in metals enforces local
charge neutrality independent of temperature. The associated variation of the
chemical potential \textit{vs}. temperature is sensitive to the energy
dependence of the density of states.

The local electron density is the integral of the electron distribution
function over all eigenstates, for a normal metal at thermal equilibrium
$n=\sum_{\mathbf{k}}f_{\mathbf{k}}$ (where $f=f^{\uparrow}=f^{\downarrow}$).
Taking into account the energy dependence of the density of states,
$\mathcal{N}(\varepsilon)$, the electron density in the limit $(k_{B}T/\mu
_{0})^{2}\ll1,$ in which we may use the Sommerfeld approximation, reads
\cite{Ashcroft:76}
\begin{equation}
n=\int d\varepsilon\mathcal{N}(\varepsilon)f_{0}(\varepsilon;\mu
_{0},T)=\mathcal{N}\left(  \mu_{0}\right)  \left(  \mu_{0}-\frac
{eS_{\mathcal{N}}T}{2}\right)  +\delta{n}(0),
\end{equation}
where $\delta{n}(0)=\int^{\mu_{0}}d\varepsilon\mathcal{N}(\varepsilon)$ is a
constant zero-temperature correction. $S_{\mathcal{N}}=-e\mathcal{L}%
_{0}T\partial_{\varepsilon}\ln\mathcal{N}(\varepsilon)|_{\mu_{0}}$ can be
interpreted as an intrinsic thermopower that does not depend on impurity
scattering. Recalling that $\sigma=e^{2}\mathcal{N}{\mathcal{D}}$ we observe a
relation with the diffuse thermopower as $S=S_{\mathcal{N}}+S_{{\mathcal{D}}}%
$. At thermal equilibrium the charge neutrality, $n(T)=n(0)$ implies
$\mu(T)=\mu_{0}+eS_{\mathcal{N}}T/2$ , \textit{i.e.}, a temperature dependent
chemical potential. The Maxwell identity $\partial\mathcal{S}/\partial
n|_{T}=-\partial\mu/\partial T|_{n}=-eS_{\mathcal{N}}/2$ \ relates
$S_{\mathcal{N}}$ to the entropy $\mathcal{S}$.

The above discussion can be extended to a ferromagnetic metal with
spin-polarized density of states. The spin-dependent electron densities vary
with temperature as
\begin{align}
n^{\left(  \alpha\right)  }(\mu,T)  &  =\int d\varepsilon\mathcal{N}^{\left(
\alpha\right)  }(\varepsilon)f_{0}(\varepsilon;\mu_{0},T)\\
&  =\mathcal{N}^{\left(  \alpha\right)  }\left(  \mu_{0}-\frac{eS_{\mathcal{N}%
}^{\left(  \alpha\right)  }T}{2}\right)  +\delta{n}^{\left(  \alpha\right)
}(0) \label{sdens}%
\end{align}
Similarly, $\delta{n}^{\left(  \alpha\right)  }(0)=\int^{\mu_{0}}%
d\varepsilon\mathcal{N}^{\left(  \alpha\right)  }(\varepsilon),$ where
$S_{\mathcal{N}}^{\left(  \alpha\right)  }=-e\mathcal{L}_{0}T\partial
_{\varepsilon}\ln\mathcal{N}^{\left(  \alpha\right)  }(\varepsilon)|_{\mu_{0}%
}$ and $\mathcal{N}^{\left(  \alpha\right)  }$ is the density of states at the
chemical potential. The transport thermopower is $S^{\left(  \alpha\right)
}=S_{\mathcal{N}}^{\left(  \alpha\right)  }+S_{{\mathcal{D}}}^{\left(
\alpha\right)  }$. At finite temperature, the spin-polarized $eS_{\mathcal{N}%
}^{(\alpha)}T/2$ act as intrinsic thermoelectric potentials, resembling the
Zeeman splitting of energy bands by an external magnetic field. The charge
neutrality condition in the normal metal holds in the ferromagnet as well,
\textit{i.e}. $\partial(n^{\left(  \uparrow\right)  }+n^{\left(
\downarrow\right)  })/\partial T=0$. Using Eq. (\ref{sdens}) this condition
implies
\begin{equation}
\frac{1}{e}\frac{\partial\mu_{0}}{\partial T}=\frac{\mathcal{N}^{\left(
\uparrow\right)  }S_{\mathcal{N}}^{\left(  \uparrow\right)  }+\mathcal{N}%
^{\left(  \downarrow\right)  }S_{\mathcal{N}}^{\left(  \downarrow\right)  }%
}{\mathcal{N}^{\left(  \uparrow\right)  }+\mathcal{N}^{\left(  \downarrow
\right)  }}=S_{\mathcal{N}}, \label{Tshift}%
\end{equation}
which reduces to that in the normal metal when $\mathcal{N}^{\left(
\uparrow\right)  }=\mathcal{N}^{\left(  \downarrow\right)  }$. A change in the
equilibrium temperature thus shifts the chemical potential (the Fermi level)
just as a gate voltage but with magnitude proportional to the intrinsic
thermopower ($\delta\mu=eS_{\mathcal{N}}\delta T$). A similar effect exists in
the presence of a uniform magnetic field, where the charge neutrality
condition implies a magnetic field-dependent chemical potential in the
ferromagnet known as the magneto-Coulomb effect
\cite{Ono:jpsj98,Shimada:prb01}. Substituting Eq. (\ref{Tshift}) into the
expressions for the spin-polarized electron densities, we find for the
variation of the spin density $n_{s}=n^{\left(  \uparrow\right)  }-n^{\left(
\downarrow\right)  }$ as a function of temperature as $\partial n_{s}/\partial
T=e\mathcal{N}S_{\mathcal{N}}(P_{\mathcal{N}}-P_{\mathcal{N}}^{\prime})$ or,
after integration over temperature,
\begin{equation}
n_{s}(T)-n_{s}(0)=\frac{e}{2}\mathcal{N}S_{\nu}T(P_{\nu}-P_{\nu}^{\prime})
\label{nsshift}%
\end{equation}
in which $P_{\mathcal{N}}$ and $P_{\mathcal{N}}^{\prime}$ are the spin
polarization of the densities of states ($\mathcal{N}^{\left(  \alpha\right)
}$) and its energy derivative ($\partial_{\varepsilon}^{\left(  \alpha\right)
}\mathcal{N}$), respectively, and $\mathcal{N}=\mathcal{N}^{\left(
\uparrow\right)  }+\mathcal{N}^{\left(  \downarrow\right)  }$. The spin
density is thus temperature dependent when $P_{S_{\mathcal{N}}}%
=(P_{\mathcal{N}}^{\prime}-P_{\mathcal{N}})/(1-P_{\mathcal{N}}^{\prime
}P_{\mathcal{N}})\neq0$.

The density of states in a ferromagnet is in general also temperature
dependent, $\partial\mathcal{N}/\partial T=\left(  \partial\mathcal{N}%
/\partial\Delta\right)  \left(  \partial\Delta/\partial T\right)  \neq0$. At
thermal equilibrium the spin density and the exchange splitting are closely
related to each other $\Delta(T)\propto n_{s}(T)$. In the Appendix we show
that even far below the Curie temperature this temperature dependence may be significant.

\section{Thermally induced spin density/polarization}

In Section 3, we derived the spin accumulation in a ferromagnetic wire under a
temperature gradient, Eq. (\ref{musx}). The chemical potential is a function
of the electron density and the temperature, as shown in the previous Section.
A local variation of the spin chemical potential from equilibrium $\delta
\mu^{\left(  \alpha\right)  }(x)=\mu^{\left(  \alpha\right)  }(x)-\mu_{0}$ can
be expanded in terms of deviations of non-equilibrium spin-polarized electron
densities and temperature ($n^{\left(  \alpha\right)  },T$) from their
equilibrium values ($n_{0}^{\left(  \alpha\right)  },T_{0}$)
\begin{equation}
\delta\mu^{\left(  \alpha\right)  }(x)=\left(  \frac{\partial\mu^{\left(
\alpha\right)  }}{\partial n^{\left(  \alpha\right)  }}\right)  _{T}\delta
n^{\left(  \alpha\right)  }(x)+\left(  \frac{\partial\mu^{\left(
\alpha\right)  }}{\partial T}\right)  _{n^{\left(  \alpha\right)  }}\delta
T(x). \label{expanddmu}%
\end{equation}
The derivatives can be found by an extension of Eq. (\ref{sdens}) by replacing
$f_{0}$ with the local spin-dependent distribution functions $f^{\left(
\alpha\right)  }(\varepsilon;\mu^{\left(  \alpha\right)  },T)$, as
\begin{align}
&  \left(  \frac{\partial\mu^{\left(  \alpha\right)  }}{\partial n^{\left(
\alpha\right)  }}\right)  _{T_{0}}=\frac{1}{\mathcal{N}^{\left(
\alpha\right)  }},\\
&  \left(  \frac{\partial\mu^{\left(  \alpha\right)  }}{\partial T}\right)
_{n_{0}^{\left(  \alpha\right)  }}=eS_{\mathcal{N}}^{\left(  \alpha\right)  }.
\end{align}
In a ferromagnet subject to a temperature gradient, the spin chemical
potential thus contains two contributions
\begin{multline}
\mu_{s}(x)=\delta n^{\left(  \uparrow\right)  }(x)/\mathcal{N}^{\left(
\uparrow\right)  }-\delta n^{\left(  \downarrow\right)  }(x)/\mathcal{N}%
^{\left(  \downarrow\right)  }\\
+e(S_{\mathcal{N}}^{\left(  \uparrow\right)  }-S_{\mathcal{N}}^{\left(
\downarrow\right)  })\left(  T(x)-T_{0}\right)
\end{multline}
In a bulk ferromagnet with spin-flip scattering we have seen that $\mu
_{s}(x)=0$, such that
\begin{multline}
\delta n^{\left(  \uparrow\right)  }(x)/\mathcal{N}^{\left(  \uparrow\right)
}-\delta n^{\left(  \downarrow\right)  }(x)/\mathcal{N}^{\left(
\downarrow\right)  }\\
=-e(S_{\mathcal{N}}^{\left(  \uparrow\right)  }-S_{\mathcal{N}}^{\left(
\downarrow\right)  })\left(  T(x)-T_{0}\right)  , \label{spinp}%
\end{multline}
which is a spin density difference (spin polarization) induced by the
temperature gradient. Eq. (\ref{spinp}), regardless of its position
dependence, is equivalent to Eq. (\ref{nsshift}) for the temperature
dependence of the spin density at thermal equilibrium. The spin polarization,
Eq. (\ref{spinp}), adopts a linear profile and changes sign in the center of
the magnetic wire, where $T_{0}=(T_{L}+T_{R})/2$. It originates from the
temperature dependence of the chemical potential in the presence of spin
polarization of the intrinsic thermopower in the ferromagnet.

The temperature dependent spin polarization and local exchange splitting in
the bulk ferromagnet is a local equilibrium effect that cannot drive a
non-equilibrium spin current into lateral contacts. The observed spin signal
in the experiments by Uchida \textit{et al}. \cite{Uchida:nat08} can therefore
not be due to the spin polarization induced by the intrinsic thermopower. A
thermally induced spin pumping mechanism due to magnon propagation in the bulk
ferromagnet \cite{Xiao:unpublished} may explain the observed spin signals.

\section{Conclusions}

We presented a theoretical analysis of thermoelectric transport in
ferromagnetic metals using an extension of the spin-dependent Boltzmann and
diffusion equations. We showed that in a bulk ferromagnetic metal subject to a
temperature gradient a pure thermal spin current is generated by spin-flip
scattering whereas the spin accumulation vanishes. The temperature dependent
spin density arising from an energy-dependent density of states generates a
spatially varying local spin polarization along the ferromagnetic wire. The
spatial dependence of the exchange splitting induced by a temperature gradient
to leading order does not lead to effects on spin transport. This work could
help to better understand thermoelectric phenomena in metallic ferromagnets.

\section*{Acknowledgments}

We thank A. Brataas, J. Ieda, P. J. Kelly, A.H. MacDonald, Y. Tserkovnyak and
J. Xiao for fruitful discussions. This work is supported by \textquotedblleft
NanoNed\textquotedblright, a nanotechnology programme of the Dutch Ministry of
Economic Affairs.

\appendix
\renewcommand{\theequation}{A-\arabic{equation}}
%redefine the command that creates the equation no.
\setcounter{equation}{0}
%reset counter

\section*{Appendix: Temperature dependence of the exchange potential in the
Stoner model}

Here we analyze the temperature dependence of the chemical potential and
equilibrium spin densities based on the Stoner theory of ferromagnetism. The
Stoner criterion $\mathcal{N}(\mu_{0})J=1,$ in which $J$ is the Stoner
exchange parameter, describes the onset of ferromagnetism. A stable magnetic
ordering leads to an exchange splitting of energy bands for opposite spin
directions, which is related to the spin density by $\Delta(T)=Jn_{s}(T)$. $J$
is an essentially atomic-like exchange integral that does not depend on
temperature. Here we study the temperature dependence of the chemical
potential and the spin density self-consistently.

We emphasize that the Stoner model is woefully inadequate at elevated
temperatures since it completely neglects spin wave excitations of the
magnetic order parameter. Consequently, the critical temperatures are
drastically overestimated by the Stoner criterion. Nevertheless we believe
that at temperatures sufficiently below the Curie transition, the Stoner
model, as a simple implementation of density-functional theory, can provide
useful qualitative insights.

The density of states depends on temperature via the exchange splitting. In
the rigid-band model $\mathcal{N}^{(\alpha)}=\mathcal{N}\left(  \varepsilon
+\alpha\Delta\right)  $. The charge neutrality condition $\partial
(n^{\uparrow}+n^{\downarrow})/\partial T=0$ now implies
\begin{equation}
\int d\varepsilon\left(  f\frac{\partial\mathcal{N}}{\partial T}%
+\mathcal{N}\frac{\partial f}{\partial T}\right)  =0
\end{equation}
where $\mathcal{N}=\mathcal{N}^{\uparrow}+\mathcal{N}^{\downarrow}$ and
$f=f^{\uparrow}=f^{\downarrow}$ at thermal equilibrium. Using the identities
\begin{equation}
\frac{\partial f}{\partial T}=\left(  -\frac{\partial f}{\partial\varepsilon
}\right)  \left(  \frac{\varepsilon-\mu}{T}+\frac{\partial\mu}{\partial
T}\right)  \neq\left(  \frac{\partial f}{\partial T}\right)  _{\mu}
\label{dfdT}%
\end{equation}
and $\partial\mathcal{N}/\partial T=\left(  \partial\Delta/\partial T\right)
\left(  \partial\mathcal{N}/\partial\Delta\right)  ,$ we find for the
temperature dependence of the chemical potential
\begin{align}
\frac{\partial\mu}{\partial T}  &  =-\chi^{-1}\int d\varepsilon\left(
\frac{\varepsilon-\mu}{T}\right)  \mathcal{N}(\varepsilon,T)\left(
-\frac{\partial f}{\partial\varepsilon}\right) \nonumber\\
&  -\chi^{-1}\frac{\partial\Delta}{\partial T}\int d\varepsilon\frac
{\partial\mathcal{N}(\varepsilon,T)}{\partial\Delta}f(\varepsilon) \label{muT}%
\end{align}
where $\chi(T)\equiv\int d\varepsilon\mathcal{N}(\varepsilon,T)\left(
-\partial_{\varepsilon}f\right)  $ is the (Pauli) susceptibility. The first
term above is the intrinsic thermopower ($=eS_{\mathcal{N}}$), and the second
term is the correction by the temperature dependence of the exchange
splitting. The above equation can be rewritten as
\begin{equation}
\frac{\partial\mu}{\partial T}=eS_{\mathcal{N}}-\frac{\chi_{s}}{\chi}%
\frac{\partial\Delta}{\partial T} \label{muT2}%
\end{equation}
in which $\chi_{s}=\chi^{\uparrow}-\chi^{\downarrow}$.

Next we are interested in the variation of spin density by temperature. The
self-consistency condition reads
\begin{align}
\frac{\partial n_{s}(T)}{\partial T}  &  =\int d\varepsilon\left(
\frac{\partial\mathcal{N}_{s}(\varepsilon,T)}{\partial T}f(\varepsilon
)+\mathcal{N}_{s}(\varepsilon,T)\frac{\partial f(\varepsilon)}{\partial
T}\right) \nonumber\\
&  =\frac{1}{J}\frac{\partial\Delta(T)}{\partial T} \label{self-cons}%
\end{align}
With Eqs. (\ref{dfdT},\ref{muT2}), the temperature dependence of the spin
density becomes
\begin{equation}
\frac{\partial n_{s}(T)}{\partial T}=\frac{e(S_{\mathcal{N}}-S_{\mathcal{N}%
_{s}})\chi_{s}}{1-J\chi\left(  1-\left(  \chi_{s}/\chi\right)  ^{2}\right)  }
\label{sd-temp}%
\end{equation}
in which $S_{\mathcal{N}_{s}}$ and $\chi_{s}$ are obtained as intrinsic
thermopower and susceptibility but with substituting $\mathcal{N}%
(\varepsilon,T)\rightarrow\mathcal{N}_{s}(\varepsilon,T)$. Equation
(\ref{sd-temp}) is exact within the limits of the validity of the rigid-band
Stoner model. Neglect of the interaction correction $\partial n_{s}/\partial
T\approx e(S_{\mathcal{N}}-S_{\mathcal{N}_{s}})\chi_{s}$ is equivalent to our
former result Eq. (\ref{nsshift}). At the Curie temperature $J\chi=1\;$and
$\partial n_{s}(T)/\partial T$ diverges. At low temperatures, assuming that
$J\chi$ does not change much from the Stoner criterion
\begin{align}
\frac{\partial n_{s}(T)}{\partial T}  &  \rightarrow\frac{e(P_{\mathcal{N}%
}-P_{\mathcal{N}}^{\prime})\chi S_{\mathcal{N}}}{1-J\chi\left(
1-P_{\mathcal{N}}^{2}\right)  }\\
&  \approx\frac{(P_{\mathcal{N}}-P_{\mathcal{N}}^{\prime})}{P_{\mathcal{N}%
}^{2}}\left(  -e\mathcal{L}_{0}T\right)  \partial_{\varepsilon}\mathcal{N}%
|_{\mu_{0}}.
\end{align}
The temperature dependence of the gap follows from Eqs. (\ref{sd-temp}%
,\ref{self-cons}), which leads to contributions to the transport thermopower
as explained below Eq. (\ref{currents2}).

\bibliographystyle{elsarticle-num}
\bibliography{pjk}

\end{document}